%% file: Saw_tooth.tex
\documentclass[aps,prl,reprint]{revtex4-1}
\usepackage{blindtext}
\usepackage{graphicx}
\usepackage[colorlinks,linkcolor=blue,anchorcolor=blue,citecolor=blue,urlcolor=blue]{hyperref}

\begin{document}
\title{Cutting and tearing thin elastic sheets: two novel single-period cracks and the first period-doubling crack}

\author{Chuang-Shi Shen \textsuperscript{1,2}}
 \author{Chun-Lin Du \textsuperscript{2}}
 \author{Huan-Fang Wang \textsuperscript{2}}
 \author{Chao Zhang \textsuperscript{2}}
  \affiliation{\textsuperscript{1} School of Civil Aviation, Zhengzhou University of Aeronautics, Zhengzhou 450015, China}%
  
 \affiliation{\textsuperscript{2} School of Aeronautics, Northwestern Polytechnical University, Xi'an 710072, China}
%


\begin{abstract}
Two novel single-period cracks were observed in experiments of cutting a folded sheet with a blunt object and tearing a thin brittle sheet under the guidance of a meterstick. Additionally, we observed a period-doubling crack in the tearing experiment. We cut and tore the sheet in different directions. The experimental results suggested that the anisotropy of the thin sheet played an important role in the formation of these two types of saw-tooth cracks. We demonstrated that the formation of the period-doubling crack was closely correlated with the changing of the contact region between the sheet and the meterstick. We also showed that the growth process of crack made by cutting was a logistic growth process (S-curve), while the cracks made by tearing propagated in the form of approximate power-law function. 
\end{abstract}

\maketitle

The classical fracture theories initially proposed by Griffith \cite{griffith1921} fall short of predicting the path of a crack as it propagates through a solid \cite{bouchbinder201401}. The study of the fracture path and the associated instabilities has been the subject of many research attempts. A body of unstable cracks has been observed in the laboratory including branching cracks and cracks with a rough surface \cite{bowden1967,ravi-chandar1984,Fineberg1991,Lubomirsky2018,bouchbinder2014}, oscillatory cracks \cite{Lubomirsky2018,bouchbinder2014,yuse1993,Ronsin1995,Deegan2001,Livne2007,chen2017,chai1986,pease2007,cai2010}, shark-fin-like cracks \cite{roman2003,Ghatak2003,Audoly2005,atkins2007,Hamm2020,Tallinen2011}, spiral cracks \cite{Leung2001,Neda2002,Marthelot2014,Fuentealba2016,Fuentealba2020}, crescent cracks \cite{Marthelot2014}, tongue-like cracks \cite{hamm2008,Bayart2011,Brau2014,Kruglova2011}, zigzag cracks \cite{wu1995,Takei2013}, en passant cracks \cite{Fender2010,Dalbe2015,Schwaab2018}, helical cracks \cite{pons2010,chen2015}, sideway cracks \cite{gent2003,lee2019}, etc. Thin sheets are ubiquitous \cite{Witten2007} such that almost everyone experienced tearing and cutting a sheet, such as opening an envelope or tearing a piece of paper in half.  Many studies have been carried out regarding these common experiences, and a number of crack growth paths were revealed \cite{roman2013}. Most previous studies assumed that a thin sheet was isotropic. However, most thin sheets are anisotropic because of their inherent micro-structure and their manufacturing process. This raises the question of whether there are any undiscovered crack propagation paths in the fracture of an anisotropic sheet. Moreover, as the previously discovered oscillatory cracks were all single-period cracks \cite{Lubomirsky2018,bouchbinder2014,yuse1993,Ronsin1995,Deegan2001,Livne2007,chen2017,chai1986,pease2007,cai2010,roman2003,Ghatak2003,Audoly2005,atkins2007,Hamm2020,Tallinen2011}, the question remained whether there are any period-doubling cracks \cite{brau2011,Losert1996}.

\begin{figure} 
	\includegraphics[scale=1]{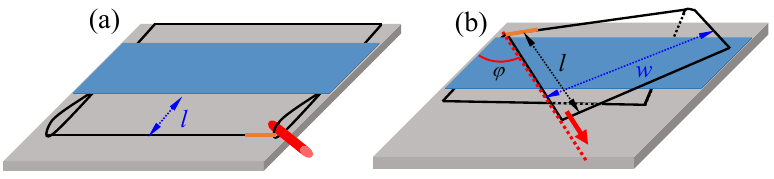}
	\caption{ (color online) Schematic of the experimental set up: (a) cutting; (b) tearing.}
	\label{fig.1}
\end{figure}

Here, we studied the cutting of a folded sheet with a blunt tool and the tearing of a sheet under the guidance of a meterstick. In both experiments, we employed anisotropic sheets of bi-axially oriented polypropylene (widely applied in fields of product packaging) of thickness $t$=$40\mu m$, $53\mu m$ with a strength that varied from 120 to 180MPa depending on the direction. All experiments were performed quasi-statically. Fig. \ref{fig.1} shows a schematic of these two experiments. After tearing and cutting, the sheets were digitized using a scanner, and the morphology of the fracture path was measured.

The first experiment [Fig. \ref{fig.1}(a)] was performed to scrutinize the crack propagation in the process of cutting a folded thin sheet with a tool, which mimicked the process of opening an envelope with a knife. The cutting process was as follows. First, a thin sheet was folded in half. No crease was formed in the sheet during the folding process. Second, this folded sheet was placed on a platform. Third, a plane was put on it and an external force was applied on the top surface of this plane, and an initial notch (orange) was cut with a knife. Finally, a blunt object (red) was utilized to cut this sheet. Two types of tools were used in terms of shape: rectangular and circular. The angle between the tool and the sheet was $\alpha {\rm{ = 3}}{{\rm{5}}^ \circ } \pm {{\rm{5}}^ \circ }$.

\begin{figure*}
	\includegraphics[scale=1]{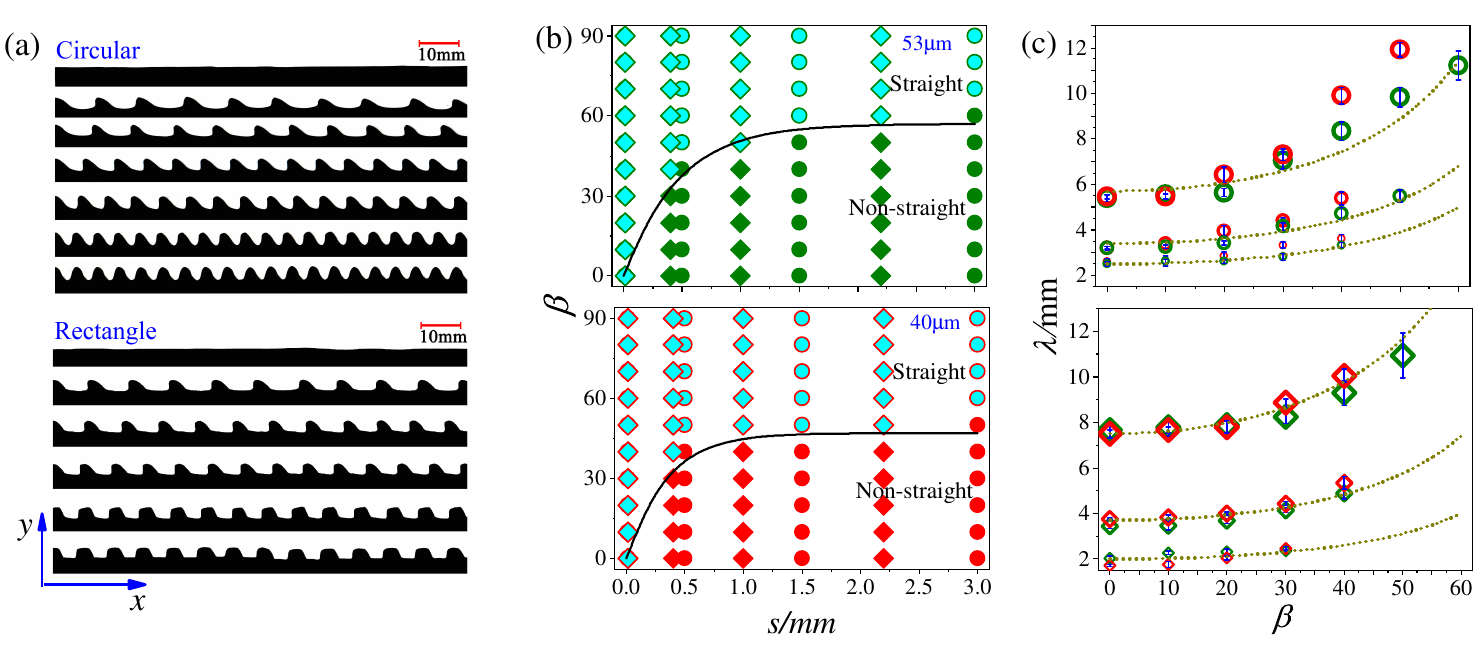}
	\caption{\label{fig.2} 
		(color online) (a) Scanned photographs of fracture path cut by a rectangular tool with a thickness of 2.2 mm (from top to bottom, $\beta  \approx {60^ \circ },{50^ \circ },{40^ \circ },{30^ \circ },{20^ \circ },{10^ \circ },{0^ \circ }$) and a circular tool with a thickness of 3 mm (from top to bottom, $\beta  \approx {50^ \circ },{40^ \circ },{30^ \circ },{20^ \circ },{10^ \circ },{0^ \circ }$ ), the sheets with thickness of 40um. (b) The phase diagram of the two kind of cracks. The line (${\beta} \propto a(1 - {e^{( - s/b)}})$, up: $a=57$, $b=0.45$; down: $a=47$, $b=0.32$), drawn to guide the eye, indicates the phase boundary between the straight and non-straight cracks. (c) The wavelength of the cracks cut by rectangle (0.4, 1, 2.2mm) and circular (0.5, 1.5, 3mm) tool as function of $\beta $. The sheets with thickness of $40\mu m$ (red) and $53\mu m$ (green). The symbols $\bigcirc$ and $\diamondsuit$ represent the circular and rectangular tool, respectively.}
\end{figure*}

\begin{figure} 
	\includegraphics[scale=1]{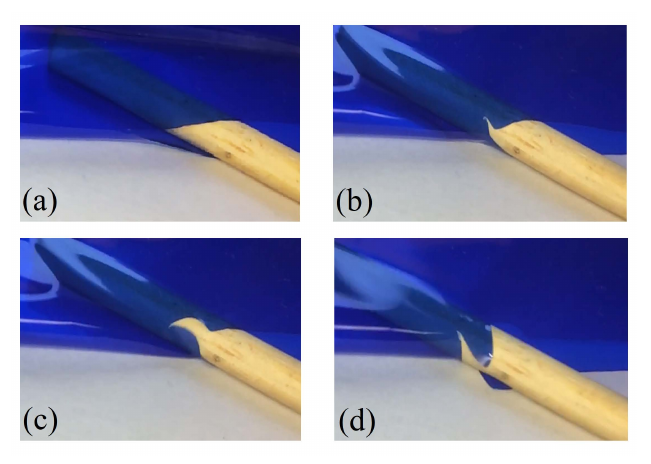}
	\caption{ (color online) Sequence of the crack propagation for $\beta  \approx {0^ \circ }$. (a) Initial symmetrical crack. The crack grows (b) along an upward inclined direction, (c) along a downward inclined direction, and (d) along an upward inclined direction.}
	\label{fig.3}
\end{figure}

Examples of the crack formed by cutting with two kinds of tools are shown in Fig. \ref{fig.2}(a). The crack path was related to the shape of the tool, which was different from the fracture path in clamped thin sheets \cite{roman2003,Ghatak2003,Audoly2005,atkins2007,Hamm2020,Tallinen2011}. The fracture path is a straight line when $\beta  > {\beta ^ * }$ (${\beta}$ was the angle between $x$-direction and the direction of the least strength of sheet). The value of  ${\beta ^ * }$ was associated with the size $s$ (thickness or diameter) of the tool. As shown in Fig. \ref{fig.2}(b), ${\beta ^ * }$ increased dramatically at the small size, but the rate of increase slowed down gradually with a further increase of the tool size. It was expected that ${\beta ^ * }$ would reach a saturation value when the thickness exceeded a certain threshold. The transition boundary between the straight crack and the non-straight crack followed ${\beta} \approx a(1 - {e^{( - s/b)}}$. For sufficiently thin tools, ${\beta ^ * } = 0$,  the crack path was a straight line, as expected.  

The amplitude (the peak-to-valley distance in the $y$-direction) of the fracture path was approximately the same as the thickness/diameter of the tool. The wavelength  $\lambda $ (the peak-to-peak distance in the $x$-direction) of the fracture path was dependent on the ${\beta}$, and $\lambda  \propto {1 \mathord{\left/{\vphantom {1 {\cos \beta }}} \right.\kern-\nulldelimiterspace} {\cos \beta }}$ [Fig.\ref{fig.2}(c)]. 

These results suggested the formation of no-straight cracks have something to do with the anisotropy of the sheets. A possible explanation is as follows. For smaller $\beta $, such as $\beta  \approx {0^ \circ }$ (the whole system was not completely symmetrical, it was difficult to ensure ${\beta ^ * } = 0$), as the tool moved toward an initial symmetrical crack [Fig. \ref{fig.3}(a), also see Supplementary Movies \cite{See}], the crack opening displacement had to be increased to accommodate the tool motion. Hence, the sheet was subjected to two forces (in-plane force ${F^I}$ and out-of-plane force ${F^O}$). Because $\sigma _{\rm{w}}^ *  < \sigma _{\rm{s}}^ * $ ($\sigma _s^ * $, $\sigma _w^ * $ denote the strength of the strong and weak direction of the sheet, respectively), the crack tended to grow along an upward-inclined direction rather than the cutting direction (perpendicular to the strongest direction) [Fig. \ref{fig.3}(b)]. The out-of-plane force applied by the tool to the upper part of the sheet was decreased to zero when the crack propagated toward the upward edge of the tool. The crack began to propagate along the cutting direction under the action of in-plane tensile force, and the out-of-plane force applied on the lower part  of the sheet showed an increasing trend during this process. When the out-of-plane force increased to a critical value, the crack began to grow along a downward-inclined direction [Fig. \ref{fig.3}(c)]. Then it reached the downward edge of the tool and ultimately propagated along the cutting direction. Then, the above processes were repeated until the cutting process was completed. However, for larger  $\beta $, such as $\beta  \approx {90^ \circ }$, although the sheet was also subjected to the out-of-plane force of the tool, the crack was not affected by this force since $\sigma _{\rm{w}}^ *  < \sigma _{\rm{s}}^ * $. The crack in this case always followed a straight path parallel to the cutting direction [see Supplementary Movies \cite{See}]. 

\begin{figure} 
	\includegraphics[scale=1]{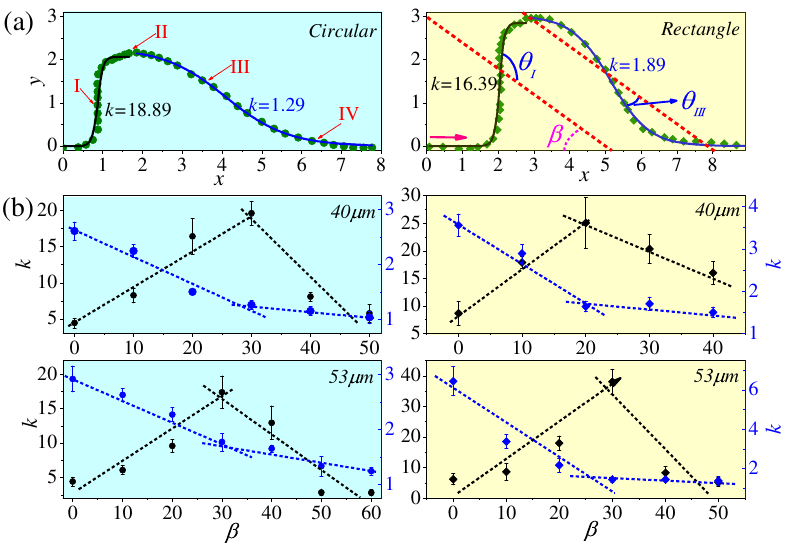}
	\caption{ (color online) (a) Details of the cutting crack path ($\beta  \approx {\rm{3}}{0^ \circ }$, $t = 40\mu m$). (b) The growth parameter $k$ for different $\beta $.}
	\label{fig.4}
\end{figure}

There were four stages in each fracture period: I, II, III, IV [Fig. \ref{fig.4}(a)]. Among them, the propagation direction of the crack in stages I and III had a certain angle with the cutting direction. For $\beta  > 0$, the crack in stage I basically propagated along the orthogonal direction of the cutting, while the crack in stage III did not. This was because ${\theta}$ (the angle between the propagation direction of crack and the direction of the least strength of sheets) in these two stages was different: ${\theta _{III}} < {90^ \circ }$ in stage III, and ${\theta _I} > {90^ \circ }$ in stage I. The absolute value of the slope $S$ of the crack in stage III was negatively correlated with $\beta $. In stages II and IV, the crack propagated along the cutting direction. However, for larger  $\beta $, the length $l$ of the crack in stage IV was significantly longer than that in stage II. This could be attributed to the different crack paths in stages I and III. The crack path in stage I was nearly perpendicular to the cutting direction, i.e. ${S_I} > {S_{III}}$. Thus, the out-of-plane force applied on the crack of stage I was larger than that on the crack of stage III, i.e. $F_I^o > F_{III}^o$, thus ${l_{II}} < {l_{IV}}$. The slope of the crack in stage I was not sensitive to $\beta $, hence, the length of the crack in stage II changed a little with $\beta $ [Fig. \ref{fig.2}(a)].

The formation process of the cutting cracks could be explained by using the logistic growth model. The formation of the non-straight crack was related to the out-of-plane force ${F^o}$. Therefore, the growth rate of the crack ${{dy}\mathord{\left/{\vphantom {{dy} {dx}}} \right.\kern-\nulldelimiterspace} {dx}}$ was positively correlated with ${F^o}$, i.e. ${{dy} \mathord{\left/{\vphantom {{dy} {dx}}} \right.\kern-\nulldelimiterspace} {dx}} \sim {F^o}$. ${F^o}$ increased initially and then diminished during the formation of the crack in stage I and stage III, as did ${{dy} \mathord{\left/{\vphantom {{dy} {dx}}} \right.\kern-\nulldelimiterspace} {dx}}$. Here we assumed ${{dy} \mathord{\left/{\vphantom {{dy} {dx}}} \right.\kern-\nulldelimiterspace} {dx}}$ followed the logistic growth model as follows:

\begin{equation}
\label{eq.1}
\frac{{dy}}{{dx}} = ky(1 - \frac{y}{s})
\end{equation}
where $k$ is the growth parameter. The size $s$ of the tool is the limit value of the function of the crack path $y$. The solution of Eq. (1) is 

\begin{equation}
\label{eq.2}
y = \frac{s}{{1 + {e^{ - k(x - {x^ * })}}}}
\end{equation}
The growth parameter $k$ should be related to $\beta $. However, because the cutting process was highly nonlinear, it was difficult for us to accurately construct a formula to describe the relationship between $k$ and $\beta $. The growth parameter $k$ was estimated by fitting Eq. (2) to experimental data. Fig. \ref{fig.4}(a) shows that the prediction of the shape of the cutting crack, $y$ (the solid line) by logistic growth model, was in good agreement with the experiments (the green scatter). For stage II to stage IV, the fitting results suggested that the parameter $k$ (absolute value) reduced with the increase of $\beta $ [Fig. \ref{fig.4}(b)], as expected.
 However, for stage IV to stage II (black), $k$ first increased and then decreased with the increase of $\beta $. The physical reasoning of this phenomenon still needs to be uncovered. 

\begin{figure*}
	\includegraphics[scale=1]{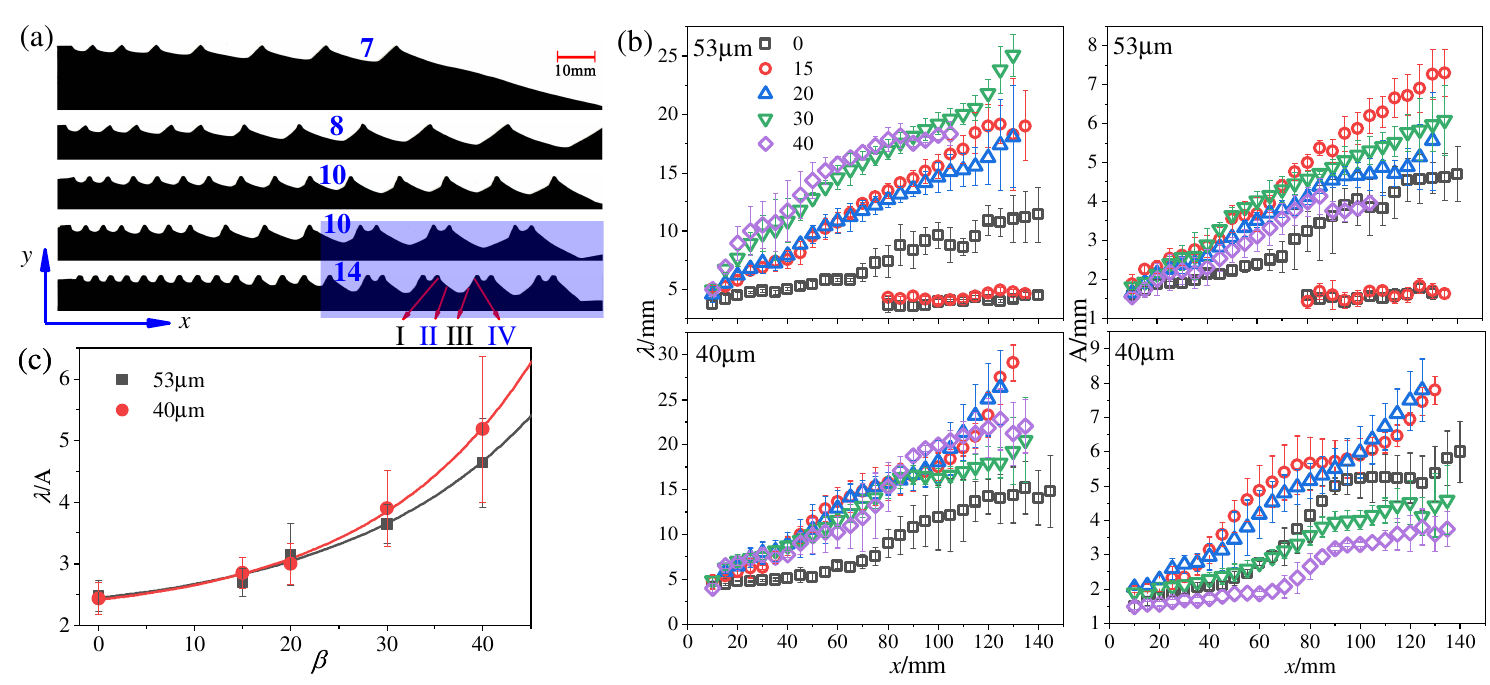}
	\caption{\label{fig.5} 
		(color online) (a)  Typical scanned photographs of the tearing cracks (From top to bottom, $\beta  \approx {40^ \circ },{30^ \circ },{20^ \circ },{15^ \circ },{0^ \circ }$),  period-doubling cracks (Light blue region), the sheet with thickness of 53um. (b) The average wavelength and the average amplitude of the crack paths. (c) The ratio of wavelength to amplitude as a function of $\beta$.}
\end{figure*}

The second experiment [Fig. \ref{fig.1}(b)] was set up to examine the crack propagation during the tearing of a sheet under the guidance of a meterstick. The tearing process was as follows: first, a thin sheet was placed on a platform. Then a meterstick  ($l$=160mm, $w$=150mm) was set on the sheet and an external force was applied on each end of this meterstick to ensure that the sheet did not slide out during the tearing process. Then an initial notch (orange) was cut with a knife. Ultimately, this sheet was torn along a line (red dashed line). The angle $\varphi $ between this line and the edge of the sheet was about ${\rm{4}}{{\rm{5}}^ \circ }$.

Fig. \ref{fig.5}(a) presents typically scanned photographs of the tearing cracks. It can be observed that the cracks formed by the tearing process were completely different from the cracks formed by cutting [Fig. 2(a)]. Depending on the angle $\beta $, the crack grew straight  ($\beta  > {55^ \circ }$) or non-straight around the meterstick. For $\beta  < {20^ \circ }$, the cracks in thick sheets ($53\mu m$) followed a period-doubling (one small and one large \cite{brau2011,Losert1996}) path when the tearing distance exceeded a critical value. In contrast, period-doubling cracks were rarely formed in thin sheets ($40\mu m$). During the last 40 years, several different types of single-period cracks have been reported, including the oscillating fracture in thermal quenching experiments \cite{yuse1993}, bi-axially stretched rubber \cite{Deegan2001}, pure uniaxial tension of thin brittle gels \cite{Deegan2001}, clamped thin sheets \cite{roman2003,Ghatak2003,Audoly2005,atkins2007,Hamm2020,Tallinen2011} and thin coatings \cite{Marthelot2014}. However, period-doubling cracks have never been reported in the literature. The cracks presented here are the \textit{\textbf{first}} period-doubling cracks.

We measured the average wavelength and amplitude of the fracture paths. As shown in Fig. \ref{fig.5}(b), the wavelength/amplitude increased during the tearing process because of the changing of the loading direction (the crack tip to the loading point). The relation between the wavelength/amplitude and $\beta $ was not monotonic function. We had no good explanation for this observation. Additional research is needed to confirm and explain this observation. However, the ratio of wavelength to amplitude increased exponentially with the increase of $\beta $ [Fig. \ref{fig.5}(c)], i.e. ${\lambda  \mathord{\left/{\vphantom {\lambda  A}} \right.\kern-\nulldelimiterspace} A} \approx c + d{e^{Rx}}$,the value of $c$, $d$ and $R$ were related to the thickness of the sheet ($40\mu m$: $c = 2.12$, $d = 0.3$ and $R = 17.15$; $53\mu m$: $c = 2.10$, $d = 0.347$ and $R = 20$).

\begin{figure} 
	\includegraphics[scale=1]{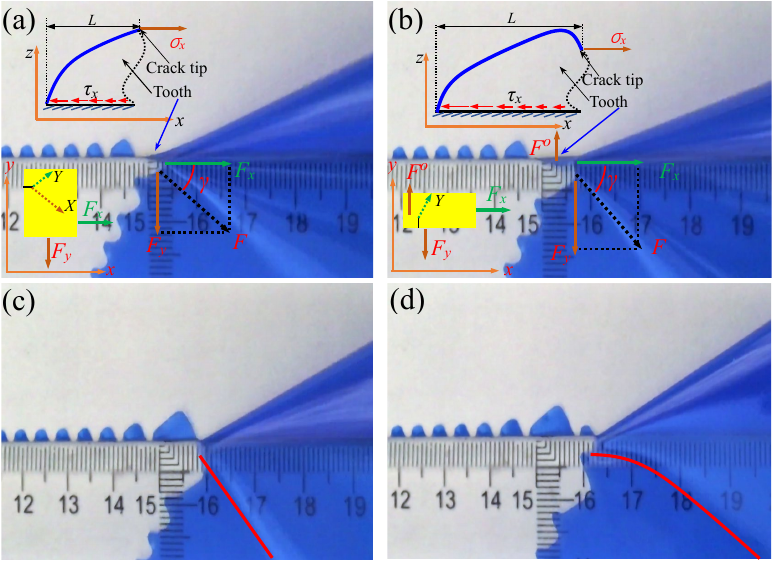}
	\caption{(color online) (a) The crack first propagated along the meterstick in a period, and (b) then propagated along the vertical direction of the meterstick. The moment when the large tooth (c) and the small tooth (d) formed}
	\label{fig.6}
\end{figure}

\begin{figure} 
	\includegraphics[scale=1]{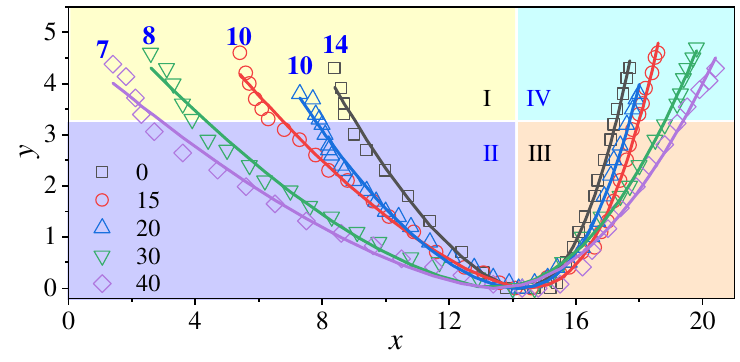}
	\caption{(color online) Details of the tearing-crack path ($t = 53\mu m$).}
	\label{fig.7}
\end{figure}

Like the cutting configuration, the formation of the crack formed by tearing was also related to the anisotropy of the material. Its formation process could also be divided into four stages: I, II, III, and IV [Fig. \ref{fig.5}(a) and Fig. \ref{fig.7}]. However, the formation process of the tearing and cutting cracks were entirely different. The tearing cracks in stage I and stage IV were formed on the side of the meterstick ($x$-$z$ plane), while the cracks in stage II and stage III were formed on the upper surface of the meterstick ($x$-$y$ plane). For smaller $\beta $, such as $\beta  \approx {{\rm{0}}^ \circ }$, the crack (mode I+II) in stage II propagated along the meterstick [Fig. \ref{fig.6}(a), also see Supplemental Movies \cite{See}] under two in-plane forces ${F_x}$ and ${F_y}$ (the crack tended to grow along X (Y) direction under ${F_x}$ (${F_y}$)). During this process, the tooth gradually became larger. As a result, the stress ${\sigma _x}$  in the crack tip gradually increased, i.e. ${\sigma _x} \propto \int_{\rm{0}}^L {{\tau _x}} dx \propto L$. When the stress ${\sigma _x}$ exceeded a certain critical value, the crack (mode I+III) then propagated (stage III) along the vertical direction of the meterstick [Fig. \ref{fig.6}(b), also see Supplemental Movies \cite{See}] due to $\sigma _{\rm{w}}^ *  < \sigma _{\rm{s}}^ * $,  mainly under in-plane force  ${F_x}$, since the out-of-plane force ${F^O}$ applied by the meter-stick to the crack tooth was small because the sheet was very thin. However, for larger $\beta $, such as $\beta  \approx {\rm{9}}{{\rm{0}}^ \circ }$, the crack always followed a straight path parallel to the meterstick for the same reason [see Supplemental Movies \cite{See}].

The shape of the tearing cracks [Fig. \ref{fig.7}] was well expressed by the following equation:
\begin{equation}
\label{eq.3}
y = \left\{ \begin{array}{l}
A{\left| {x - {x^ * }} \right|^{{p_{_l}}}}\begin{array}{*{20}{c}}
{}&{x < {x^ * }}
\end{array}\\
A{\left| {x - {x^ * }} \right|^{{p_{_{r}}}}}\begin{array}{*{20}{c}}
{}&{x > {x^ * }}
\end{array}
\end{array} \right.
\end{equation}
where ${x^ * }$ is the valley of the tearing crack, which indicates that the tearing cracks grow in the form of power-law \cite{west2017}, rather than S-curve.

To understand why the period-doubling cracks were formed, we recorded the formation process of the tearing cracks [see Supplemental Movies \cite{See}]. The deformation morphology of the sheet when the large tooth and the small tooth had just been formed is shown in Fig. \ref{fig.6}(c) and \ref{fig.6}(d), respectively. The red line is a boundary. On its left side, the sheet was attached to the meterstick. On its right side, we found that the boundaries in Fig. \ref{fig.6}(c) and \ref{fig.6}(d) were very different. In Fig. \ref{fig.6}(c), the boundary was a straight line. Conversely, the boundary was a curve in Fig. \ref{fig.6}(d). We watched the recorded video repeatedly. We found that the straight boundary did not appear in the formation process of the single-period cracks. It seemed that the straight boundary appeared only after the formation of the large tooth. A small tooth was always formed after its appearance.

In conclusion, we have reported two new period-single cracks and a period-doubling crack in cutting and tearing thin elastic sheets. We showed that they stemmed from the anisotropy of the sheet and the interaction between the sheet and the tool or meterstick. The experimental results of the present study can be a motivation for the development of new theoretical models to accurately predict the growth process of these types of cracks and to understand how the anisotropy of material affects the fracture path \cite{Hakim2005,teichtmeister2017,li2019}. These robust cracks can be employed as good test cases for theoretical models that couple anisotropy of material and fracture.

\bibliographystyle{apsrev4-1} 
\input{Saw_tooth.bbl}

\end{document}

%% file: Saw_tooth.bbl
%